\documentclass[conference]{IEEEtran}
\IEEEoverridecommandlockouts
\usepackage{cite}
\usepackage{amsmath,amssymb,amsfonts}
\usepackage{algorithmic}
\usepackage{graphicx}
\usepackage{textcomp}
\usepackage{xcolor}
\usepackage{csvsimple}
\usepackage{booktabs}
\usepackage{algorithm,algorithmic}
\usepackage{breqn}
\usepackage{subcaption}

\def\BibTeX{{\rm B\kern-.05em{\sc i\kern-.025em b}\kern-.08em
    T\kern-.1667em\lower.7ex\hbox{E}\kern-.125emX}}
\begin{document}

\title{Exploring market power using deep reinforcement learning for intelligent bidding strategies
\thanks{EPSRC}
}

\author{\IEEEauthorblockN{Alexander J. M. Kell\textsuperscript{\textsection}}
\IEEEauthorblockA{\textit{Department of Chemical Engineering} \\
\textit{Imperial College London}\\
London, U.K. \\
a.kell@imperial.ac.uk}
\and
\IEEEauthorblockN{Matthew Forshaw, A. Stephen McGough}
\IEEEauthorblockA{\textit{School of Computing} \\
\textit{Newcastle University}\\
Newcastle \\
\{matthew.forshaw, stephen.mcgough\}@newcastle.ac.uk}
}

\maketitle
\begingroup\renewcommand\thefootnote{\textsection}
\footnotetext{This work was carried out whilst at Newcastle University.}
\endgroup

\begin{abstract}


Decentralized electricity markets are often dominated by a small set of generator companies who control the majority of the capacity. In this paper, we explore the effect of the total controlled electricity capacity by a single, or group, of generator companies can have on the average electricity price. We demonstrate this through the use of ElecSim, a simulation of a country-wide energy market. We develop a strategic agent, representing a generation company, which uses a deep deterministic policy gradient reinforcement learning algorithm to bid in a uniform pricing electricity market. A uniform pricing market is one where all players are paid the highest accepted price. ElecSim is parameterized to the United Kingdom for the year 2018. This work can help inform policy on how to best regulate a market to ensure that the price of electricity remains competitive.

We find that capacity has an impact on the average electricity price in a single year. If any single generator company, or a collaborating group of generator companies, control more than ${\sim}$11$\%$ of generation capacity and bid strategically, prices begin to increase by ${\sim}$25$\%$. The value of ${\sim}$25\% and ${\sim}$11\% may vary between market structures and countries. For instance, different load profiles may favour a particular type of generator or a different distribution of generation capacity. Once the capacity controlled by a generator company, which bids strategically, is higher than ${\sim}$35\%, prices increase exponentially. We observe that the use of a market cap of approximately double the average market price has the effect of significantly decreasing this effect and maintaining a competitive market. A fair and competitive electricity market provides value to consumers and enables a more competitive economy through the utilisation of electricity by both industry and consumers.



\end{abstract}


\section{Introduction}
\label{sec:introduction}

In this paper, we use a monte-carlo simulation to model a decentralized electricity market. Decentralized electricity markets have a number of different market structures: day-ahead, balancing market and reserve markets. The day-ahead market settles payments for electricity for the following day. The balancing market is designed to regulate the frequency of the market through matching supply and demand \cite{Kell2020c}. Whereas, reserve markets provide capacity in case there are fluctuations in supply and demand, and electricity demand can't be matched with supply. Our model simulates a day-ahead market over a single year time period to explore the effects of market power on electricity prices. For instance, in varying fuel prices within a year, and generator costs based upon geographical and factors that occur by chance, such as breakdowns.

Under perfectly competitive electricity markets, generator companies (GenCos) tend to bid their short-run marginal costs (SRMC) when bidding into the day-ahead electricity market. SRMC is the cost to produce a single MWh of electricity and excludes capital costs. However, electricity markets are often oligopolistic, where a small subset of GenCos provide the majority of the capacity to the market. Under these conditions, it is possible that the assumption that GenCos are price-takers does not hold. That is, large GenCos artificially increase the price of electricity to gain an increased profit using their market power. If they were price-takers, they would have to accept the competitive price set by the market.

Reduced competition within electricity markets can lead to higher prices to consumers, with no societal benefit. It is, therefore, within the interests of the consumer and that of government to maintain a competitive market. Low energy costs enable innovation in other industries reliant on electricity, and in turn, make for a more productive economy. In this paper, we explore how to increase competition within electricity markets.

We explore the effect of total control over capacity on electricity prices. Specifically, we model different sizes of GenCos and groups of colluding GenCos, bidding strategically to maximize their profit using a learning algorithm. This is in contrast to the strategy of bidding using the SRMC of their respective power plants, which would occur under perfect market competition. Perfect market competition is a theoretical market structure where all firms sell an identical product, all firms are price-takers and market share has no influence on price. To model this we use an application of deep reinforcement learning (RL) to calculate a bidding strategy for GenCos in a day-ahead market. These GenCos are modelled as agents within the agent-based model, ElecSim \cite{Kell, Kell2020}. We use the UK electricity market instantiated on 2018 as a case study, similar to our work in \cite{Kell2019a}. That is, we model each GenCo with their respective power plants in the year 2018 to 2019. In total, we model 60 GenCos with 1,085 power plants. It is possible, however, to generalise this approach and model to any other decentralized electricity market. 

We use the deep deterministic policy gradient (DDPG) deep RL algorithm, which allows for a continuous action space \cite{Hunt2016a}. Conventional RL methods require discretization of state or action spaces and therefore suffer from the curse of dimensionality \cite{Ye2020a}. As the number of discrete states and actions increases, the computational cost grows exponentially. However, too small a number of discrete states and actions will reduce the information available to the GenCos, leading to sub-optimal bidding strategies. Additionally, by using a continuous approach, we allow for GenCos to consider increasingly complex bidding strategies.

The deep RL algorithm uses big data from the simulation to learn an optimal bidding strategy. Specifically, it uses the reward, action taken and state space to converge to an optimal behaviour. Over the course of a single training session, the deep RL algorithm observes data between 600,000 and 800,000 times in the form of steps and takes actions during each  step. Each state space contains six data points.

Additionally, the simulation model, ElecSim, is parametrized by real-world data. This real-world data includes costs of power plants, solar and wind conditions, electricity demand, company finances, historical carbon and fuel prices, amongst others. Using this data, rules are used to generate data used by the simulation.

Other works have considered a simplified model of an electricity market by modelling a small number of GenCos or plants \cite{EsmaeiliAliabadi2017,Tellidou2007}. We, however, model each GenCo as per the UK electricity market with their respective power plants in a day-ahead market. In addition, further work focuses on a bidding strategy to maximize profit for a GenCo. However, in our work, we focus on the impact that large GenCos, or colluding groups of GenCos, can have on electricity price.

Our approach does not require GenCos to formulate any knowledge of the data which informs the market-clearing algorithm or rival GenCo bidding strategies, unlike in game-theoretic approaches \cite{Wang2011}. This enables a more realistic simulation where the strategy of rival GenCos are unknown.

The contributions of this work are to use a novel agent-based model to explore market power in electricity markets using a deep reinforcement learning algorithm. Oligopolistic markets often exhibit market power which can lead to artificially inflated electricity prices. We find an optimal market cap level in the UK electricity market as well as the maximum proportion of capacity that a single entity should own to prevent imperfect market conditions.

In Section \ref{sec:lit-review} we review the literature, and explore other uses of RL in electricity markets. In Section \ref{sec:material} we introduce the agent-based model used and the DDPG algorithm. Section \ref{sec:methodology} explores the methodology taken for our case study. The results are presented in Section \ref{sec:results}. We discuss and conclude our work in Sections \ref{sec:discussion} and \ref{sec:conclusion} respectively.








\section{Literature Review}
\label{sec:lit-review}

Intelligent bidding strategies for day-ahead electricity markets can be divided into two broad categories: game-theoretic models and simulation. Game-theoretic approaches may struggle in complex electricity markets where Nash equilibriums do not exist \cite{Wang2011}. Agent-based models (ABMs) allow for the simulation of heterogenous irrational actors with imperfect information. Additionally, ABMs allow for learning and adaption within a dynamic environment \cite{EsmaeiliAliabadi2017}.

\subsection{Game-theoretic approaches}

Here, we explore game-theoretic approaches. Kumar \textit{et al.} propose a Shuffled Frog Leaping Algorithm (SFLA) \cite{VijayaKumar2014} to find bidding strategies for GenCos in electricity markets. SFLA is a meta-heuristic that is based on the evolution of memes carried by active individuals, as well as a global exchange of information among the frog population. They test the effectiveness of the SFLA algorithm on an IEEE 30-bus system and a practical 75-bus Indian system. A bus in a power system is a vertical line in which several components are connected in a power system. For example, generators, loads, and feeders can all be connected to a bus. Kumar \textit{et al.} find superior results when compared to particle swarm optimization and the genetic algorithm with respect to total profit and convergence with CPU time. They assume that each GenCo bids a linear supply function, and model the expectation of bids from rivals as a joint normal distribution. In contrast to their work, we do not require an estimation of the rival bids.

Wang \textit{et al.} propose an evolutionary imperfect information game approach to analyzing bidding strategies with price-elastic demand \cite{Wang2011}. Their evolutionary approach allows for GenCos to adapt and update their beliefs about an opponents' bidding strategy during the simulation. They model a 2-bus system with three GenCos. Our work, however, models a simulation with 60 GenCos across the entire UK, which would require a 28-bus system model \cite{Bell2010}.

\subsection{Simulation}

Simulations are a software system which can replicate the salient features of a real-world system. Through the modelling of such systems using simulation, `what-if' analysis can be undertaken. These `what-if' analysis can be used to gain a greater understanding how a system would behave under a change in environment or policy. Simulation is especially useful when performing changes to the real system would be unpalatable to perform - such as having a significant impact or having a high cost. 

`Digital twins' are a specific instance of a simulation. For example, a simulation can be built of a decentralized electricity market, whereas a digital twin would be a simulation of the UK electricity market. A digital twin allows for rapid experimentation to be carried out, in faster than real-time, and avoid the risk that this experimentation could have. 

Simulations are often used to understand markets and how an individual or entity should optimally behave. This is due to the fact that there is a high risk involved in submitting bids due to the potential negative effects of having a bid rejected. In addition to this, the uncertainties in a market can be better understood through monte-carlo simulation, by exploring various scenarios.

\subsection{Reinforcement learning to model intelligence}

Simulation is often used as an environment for RL to learn and act in. This is due to the ability for simulation to perform actions in faster than real-time. In this section we explore reinforcement learning approaches used to make intelligent bidding decisions in electricity markets. RL is a suitable method for analyzing the dynamic behaviour of complex systems with uncertainties. RL can, therefore be used to identify optimal bidding strategies in energy markets \cite{Yang2020}. In the following papers simulations are used as the environment.

Aliabadi \textit{et al.} utilize an ABM and the Q-learning algorithm to study the impact of learning and risk aversion on GenCos in an oligopolistic electricity market with five GenCos \cite{EsmaeiliAliabadi2017}. They find that some level of risk aversion is beneficial, however excessive risk degrades profits by causing an intense price competition. Our paper focuses on the impact of the interaction of 60 GenCos within the UK electricity market. In addition, we extend the Q-learning algorithm to use the DDPG algorithm, which uses a continuous action space. A continuous action space allows for more granular action spaces to be taken, allowing for a more precise action. In addition, by modelling 60 GenCos, we are able to more closely match the real-life dynamics of the UK electricity market. 

Bertrand \textit{et al.} use RL in an intraday market \cite{Bertrand2019}. Specifically, they use the REINFORCE algorithm to optimize the choice of price thresholds. The REINFORCE algorithm is a gradient-based method. They demonstrate an ability to outperform the traditionally used method, the rolling intrinsic method, by increasing profit per day by 4.2\%. The rolling intrinsic method accepts any trade, which gives a positive profit if the contracted quantity remains in the bounds of capacity. In our paper, we model a day-ahead market and use a continuous action for price bids.

Ye \textit{et al.} propose a novel deep RL based methodology which combines the DDPG algorithm with a prioritized experience replay (PER) strategy \cite{Ye2020a}. The PER samples from previous experience, but samples from the ``important'' ones more often \cite{Schaul2016}. The PER is a modification of the often used experience buffer, which is a buffer which stores previous transitions and samples uniformly. This helps to reduce the correlations between recent experiences. They use a day-ahead market with hourly resolution and show that they are able to achieve approximately 41\%, 20\% and 11\% higher profit for the GenCo than the MPEC, Q-learning and DQN methods, respectively. In our paper, we instead look at how to prevent GenCos (or sets of colluding GenCos) from forcing higher prices above market rates.

Zhao \textit{et al.} propose a modified RL method, known as the gradient descent continuous Actor-Critic (GDCAC) algorithm \cite{Zhao2016}. This algorithm is used in a double-sided day-ahead electricity market simulation. Where in this case, a double-sided day-ahead market refers to GenCos selling their supply to distribution companies, retailers or large consumers. Their approach performs better in terms of participant's profit or social welfare compared with traditional table-based RL methods, such as Q-Learning. Our work also looks at improving on table-based methods by using function approximators.

\section{Methodology}
\label{sec:material}

In this section, we describe the RL methodology used for the intelligent bidding process as well as the simulation model used as the environment.

\subsection{Reinforcement Learning background}


In RL an agent interacts with an environment to maximize its cumulative reward. RL can be described as a Markov Decision Process (MDP). An MDP includes a state-space $\mathcal{S}$, action space $\mathcal{A}$, a transition dynamics distribution $p(s_{t+1}|s_t,a_t)$ and a reward function, where $r:S\times \mathcal{A} \rightarrow \mathbb{R}$. At each time step an agent receives an observation of the current state which is used to modify the agent's behaviour.

An agent's behaviour is defined by a policy, $\pi$. $\pi$ maps states to a probability distribution over the actions $\pi:\mathcal{S}\rightarrow \mathcal{P}(\mathcal{A})$. The return from a state is defined as the sum of discounted future reward $R_t=\sum_{i=t}^T\gamma^{(i-t)}r(s_i,a_i)$. Where $\gamma$ is a discounting factor $\gamma \in [0,1]$. The return is dependent on the action chosen, which is dependent on the policy $\pi$. The goal in reinforcement learning is to learn a policy that maximizes the expected return from the start distribution $J=\mathbb{E}_{r_i,s_i \sim E,a_i \sim \pi}[R_1]$. 

The expected return after taking an action $a_t$ in state $s_t$ after following policy $\pi$ can be found by the action-value function. The action-value function is used in many reinforcement learning algorithms and is defined in Equation \ref{eq:action-value}.
\begin{equation}
	\label{eq:action-value}
	Q^{\pi}(s_t,a_t)=\mathbb{E}_{r_{i\geq t},s_{i>t}\sim \mathcal{E},a_{i>t}\sim\pi}[R_t|s_t,a_t].
\end{equation}
\noindent The action-value function defines the expected reward at time $t$, given a state $s_t$ and action $a_t$ when under the policy $\pi$.

\subsection{Q-Learning}

 An optimal policy can be derived from the optimal $Q$-values $Q_*(s_t,a_t)=\max_\pi Q_\pi(s_t,a_t)$ by selecting the action corresponding to the highest Q-value in each state.

Many approaches in reinforcement learning use the recursive relationship known as the Bellman equation, as defined in Equation \ref{eq:bellman}:
\begin{dmath}
	\label{eq:bellman}
	Q^\pi(s_t,a_t)=\mathbb{E}_{{r_t},s_{t+1}\sim E} [r(s_t,a_t)+
	\gamma\mathbb{E}_{a_{t+1}\sim \pi}[Q_\pi(s_{t+1},\pi(s_{t+1}))]].
\end{dmath}
\noindent The Bellman equation is equal to the action which maximizes the reward plus the discount factor multiplied by the next state's value, by taking the action after following the policy in state $s_{t+1}$ or $\pi(s_{t+1})$.

The Q-value can therefore be improved by bootstrapping. This is where the current value of the estimate of $Q_\pi$ is used to improve its future estimate, using the known $r(s_t,a_t)$ value. This is the foundation of Q-learning \cite{Gay2007}, a form of \textit{temporal difference} (TD) learning \cite{Sutton2015}, where the update of the Q-value after taking action $a_t$ in state $s_t$ and observing reward $r_t$, which results in state $s_{t+1}$ is:
\begin{equation}
	Q(s_t,a_t)\leftarrow Q(s_t,a_t)+\alpha\delta_t,
\end{equation}
\noindent where,
\begin{equation}
	\delta_t=r_t+\gamma\max_{a_{t+1}}Q(s_{t+1},a_{t+1})-Q(s_{t},a_t),
\end{equation}
\noindent $\alpha\in [0,1]$ is the step size, $\delta_t$ represents the correction for the estimation of the Q-value function and $r_t+\gamma\max_{a_{t+1}}Q(s_{t+1},a_{t+1})$ represents the target Q-value at time step $t$.

It has been proven that if the Q-value for each state action pair is visited infinitely often, the learning rate $\alpha$ decreases over time step $t$, then as $t\rightarrow \infty$, $Q(s,a)$ converges to the optimal $Q_*(s,a)$ for every state-action pair \cite{Gay2007}.

However, Q-learning often suffers from the curse of dimensionality, because the Q-value function is stored in a look-up table which therefore requires the action and state spaces to be discretized. As the number of discretized states and actions increases, the computational cost increases exponentially, making the problem intractable. 

\subsection{Deep Deterministic Gradient Policy}

Many problems are naturally discretized which are well suited to a Q-learning approach, however this is not always the case, such as the problem in this paper. It is not straightforward to apply Q-learning to continuous action spaces. This is because in continuous spaces, finding the greedy policy requires an optimization of $a_t$ at every time step. Optimizing for $a_t$ at every time step would be too slow to be practical with large, unconstrained function approximators and nontrivial action spaces \cite{Hunt2016a}. To solve this, an actor-critic approach based on the deterministic policy gradient (DPG) algorithm is used \cite{Silver2014}. This approach has not often been taken in the literature, and therefore is a novelty of our approach.

The DPG algorithm maintains a parameterized actor function $\mu(s|\theta^\mu)$ which specifies the current policy by deterministically mapping states to a specific action. The critic $Q(s,a)$ is learned using the Bellman equation, as in Q-learning. The actor is updated by applying the chain rule to the expected return from the start distribution $J$ with respect to the actor parameters:
\begin{align}
\begin{split}
	\triangledown_{\theta^\mu}J\approx\mathbb{E}_{s_t\sim\rho^\beta}[\triangledown_{\theta^\mu}Q(s,a|\theta^Q)|_{s=s_t,a=\mu(s_t|\theta^\mu)}] \\
	= \mathbb{E}_{s_t\sim\rho^\beta}[\triangledown_aQ(s,a|\theta^Q)|_{s=s_t,a=\mu(s_t)}\triangledown_{\theta_\mu}\mu(s|\theta^\mu)|_{s=s_t}].
\end{split}
\end{align}
 
This is the policy gradient, the gradient of the policy's performance. The policy gradient method optimizes the policy directly by updating the weights, $\theta$, in such a way that an optimal policy is found within finite time. This is achieved by performing gradient ascent on the policy and its parameters $\pi^\theta$.

Introducing non-linear function approximators, however, means that convergence is no longer guaranteed. Although these function approximators are required in order to learn and generalize on large state spaces. The Deep Deterministic Gradient Policy (DDPG) modifies the DPG algorithm by using neural network function approximators to learn large state and action spaces online.

A replay buffer is utilized in the DDPG algorithm to address the issue of ensuring that samples are independently and identically distributed. The replay buffer is a finite-sized cache, $\mathcal{R}$. Transitions are sampled from the environment through the use of the exploration policy, and the tuple $(s_t,a_t,r_t,s_{t+1})$ is stored within this replay buffer. $\mathcal{R}$ discards older experiences as the replay buffer becomes full. The actor and critic are trained by sampling from $\mathcal{R}$ uniformly. 

A copy is made of the actor and critic networks, $Q'(s,a|\theta^{Q'})$ and $\mu'(s|\theta^{\mu'})$ respectively. These are used for calculating the target values. To ensure stability, the weights of these target networks are updated by slowly tracking the learned networks. Pseudo-code of the DDPG algorithm is presented in Algorithm \ref{alg:ddpg}.

\begin{algorithm}
\caption{DDPG Algorithm \cite{Hunt2016a}}
\begin{algorithmic}[1]
  \small
  \STATE Initialize critic network $Q(s,a|\theta^Q)$ and actor $\mu(s|\theta^\mu)$ with random weights $\theta^Q$ and $\theta^\mu$
  \STATE Initialize target network $Q'$ and $\mu'$ with weights $\theta^{Q'}\leftarrow\theta^Q,\theta^{\mu'}\leftarrow \theta^{\mu}$
  \STATE Initialize replay buffer $R$
  \FOR{\texttt{episode=1,M}}
        \STATE Initialize a random process $\mathcal{N}$ for action exploration
        \STATE Receive initial observation state $s_1$
        \FOR{\texttt{t=1,T}}
        	\STATE Select action $a_t=\mu(s_t|\theta^{\mu})+\mathcal{N}_t$ according to the policy and exploration noise, $\mathcal{N}_t$
        	\STATE Execute action $a_t$ and observe reward $r_t$ and new state $s_{t+1}$
        	\STATE Store transition $(s_t, a_t, r_t, s_{t+1})$ in $R$
        	\STATE Sample a random minibatch of $N$ transitions $(s_i, a_i, r_i, s_{i+1})$ from $R$
        	\STATE Set $y_i=r_i+\gamma Q'(s_{i+1},\mu'(s_{i+1},\mu'(s_{i+1}|\theta^{\mu'})|\theta^{Q'})$
        	\STATE Update critic by minimizing the loss: $$L=\frac{1}{N}\sum_i(y_i-Q(s_i,a_i|\theta^Q))^2$$
        	\STATE Update the actor policy using the sampled policy gradient: $$\triangledown_{\theta^\mu}J\approx \frac{1}{N}\sum_i\triangledown_a Q(s,a|\theta^Q)|_{s=s_i,a=\mu(s_i)}\triangledown_{\theta^\mu}\mu(s|\theta^\mu)|_{s_i}$$
        	\STATE Update the target networks:
        	$$\theta^{Q'}\leftarrow\tau\theta^Q+(1-\tau)\theta^{Q'}$$
        	$$\theta^{\mu'}\leftarrow\tau\theta^\mu+(1-\tau)\theta^{\mu'}$$
        \ENDFOR
      \ENDFOR
\end{algorithmic}
\label{alg:ddpg}
\end{algorithm}

\subsection{Simulation}

We utilized the long-term electricity market agent-based model, ElecSim \cite{Kell,Kell2020}. ElecSim is a simulation which consists of multiple generation companies which interact in a simulated electricity market. This electricity market is able to run for 20+ years and can show what long-term scenarios may develop from certain scenario configurations, such as high carbon and fuel prices. The generation companies invest in power plants based upon their understanding and expectations of market conditions now and in the future. They then sell the electricity generators on a day-ahead spot market. The model was run using a short term approach by only iterating through a single year (2018), composed of eight representative days, each of 24 time steps.

ElecSim is made up of six components: 1) power plant data; 2) scenario data; 3) the time steps of the algorithm; 4) the power exchange; 5) the investment algorithm and 6) the generation companies (GenCos) as agents. For this paper, we ignore the investment algorithm, due to investments happening only once a year, and not in the first year of operation. 

The power plant data is made up of both historical power plant costs and modern power plant costs, as well as running capacity limits that these plants can run. The scenario data links to many different sources, such as electricity demand, power plants in operation and fuel costs. At each time-step, a large amount of data is generated and analysed. For instance, 1085 power plants must make bids into each of the 20 demand segments, for each hour of each day, with eight days per year. This leads to over 4,000,000 actions per simulation run (1085 power plants multiplied by 20 demand segments, multiplied by 24 hours, multiplied by eight days).

The power exchange then must sort each of these bids per demand segment, from lowest bid to highest, accepting bids until the electricity demand is met. Once the electricity demand is met, the rest of the bids are rejected. There are 60 GenCos, modelled on the real-life GenCos in the UK electricity market. Each of these GenCos must then make decisions based on investments in the future, based upon the money available to them, and regressions based on expected scenarios in the future, such as carbon, fuel price and demand. In this paper, the model was only run for a single year. 

ElecSim uses a subset of representative days of electricity demand, solar irradiance and wind speed to approximate a full year. Representative days, in this context, are a subset of days which can closely approximate an entire year's electricity demand and weather patterns. By using a subset, we are able to reduce the computational time to run, whilst maintaining accuracy of modelling the real electricity market \cite{Kell2020}. We used eight representative days with an hourly granularity. This leads to 192 time-steps per year. If we did not use representative days, the simulation would consist of 365 days of 24 hour time-steps, leading to 8,760 time-steps. An increase of ${\sim}$45$\times$. A single simulation takes two minutes to run using eight representative days, which would increase to 90 minutes if representative days weren't used. This would significantly reduce the tractability of the problem, as 1,000s of simulation iterations would be required. A problem which currently took a week to run, would take 45 weeks to finalise, without a significant improvement in accuracy.

\begin{figure}
    \includegraphics[width=0.49\textwidth]{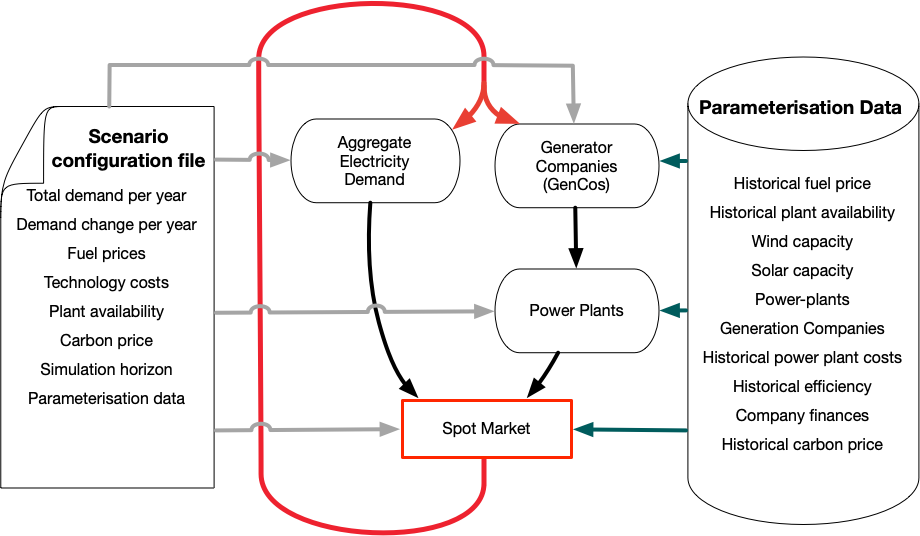}
    \caption{System overview of ElecSim \cite{Kell}.}
    \label{fig:model_details}
\end{figure}

Figure \ref{fig:model_details} shows how the six components interact. The electricity demand is matched with the supply of the power plants, owned by the generator companies (GenCos). We have configured ElecSim to model the UK electricity market, using the configuration file. Specifically, we model the actual GenCos with their respective power plants that were in operation in 2018.

The market has a uniform pricing bidding mechanism. A uniform pricing mechanism is one where a single price is paid for all electrical capacity, irrespective of the bid. The bid that is paid is that of the highest accepted price. This incentivizes GenCos to bid their SRMC. SRMC is the price that it takes to generate a single MWh of electricity, excluding capital costs. 

In this work we explore whether large GenCos, or group of GenCos, can manipulate the price of the electricity market through virtue of their size. We achieve this by allowing a subset of GenCos to bid away from their SRMC and allow them to learn an optimal bidding strategy for maximizing their income. The GenCo agents adopt a DDPG RL algorithm to select their bids. This is to explore whether large GenCos, or a group of GenCos can manipulate the price of the electricity market through market power. The remaining GenCos, which fall outside of this group, maintain a bidding strategy based upon their SRMC.

For the purpose of this work, we do not consider flow constraints within the electricity mix. This is because we model the entire UK with 1,000+ generators, and many nodes and buses. This would make the optimization problem intractable for the purpose of our simulation, especially when considering the many episodes required for training. It takes ${\sim}125$ seconds to run each episode (representing a single year in the simulation) with our current setup. By increasing the simulation time further, we would make the compute time intractable. Additionally, we must train several reinforcement learning policies to account for each GenCo and market cap. Therefore a simulation that takes ${\sim}$2 minutes to run, repeated 1,000 times takes ${\sim}$33 hours in total to finish, multiplied by 12 different GenCos takes ${\sim}$16 days, and with two market cap scenarios it would take ${\sim}$1 month.




\section{Experimental Setup}
\label{sec:methodology}

\subsection{Model Parametrization}

To parameterize the simulation, we use data from the United Kingdom in 2018. This included 1085 electricity generators and power plants with their respective GenCos. The data for this was taken from the BEIS DUKES dataset \cite{dukes_511}. The data included construction costs, capital and operating costs as well as lifetimes of the plant and capacity. The storage size of all of these power plants was 10MB of data. The electricity load data was modelled using data from \cite{gbnationalgridstatus_2019}; offshore and onshore wind and solar irradiance data from \cite{Pfenninger2016} and are 20MB in size. It would be possible to adopt this approach to other decentralized markets in other countries.

\subsection{Reinforcement Learning Setup}

Much of the data used by the algorithm was generated by the simulation and persisted in memory. With over 4,000,000 bids made by all of the power plants in each simulation run. Each bid had six different state data and a reward; leading to 28,000,000 data points per simulation. 1,000 simulation runs, therefore leads to $28\times 10^9$ data points used to train the reinforcement learning algorithm.

By modelling bidding decisions as a RL algorithm, we hope to observe the ability for RL to find the point at which market power artificially inflates electricity prices. To achieve this, we choose the six largest GenCos in the UK, as well as a smaller GenCo as a control. Groups of GenCos are modelled as a single GenCo with a single RL strategy for the purpose of this paper. Table \ref{table:genco_table} displays the groups of GenCos, as well as individual GenCos, with their respective capacity and number of plants.

\begin{table*}
\renewcommand{\arraystretch}{1.35}
\centering
\begin{tabular}{llllll}
\toprule
GenCo Groups                                                                                                                  & Capacity (MW) & Num. of Plants \\ \midrule
Orsted                                                                                                                       & 2738.7   & 11               \\
Drax Power Ltd                                                                                                               & 4035.0   & 3                \\
Scottish power                                                                                                               & 4471.5   & 49               \\
Uniper UK Limited                                                                                                            & 6605.0   & 9                \\
SSE                                                                                                                          & 8390.7   & 130              \\
RWE Generation SE                                                                                                            & 8664.0   & 11               \\
EDF Energy                                                                                                                   & 14763.0  & 14               \\
\{EDF Energy, RWE Generation SE\}                                                                                              & 23427.0  & 25               \\
\{EDF Energy, RWE Generation SE, SSE\}                                          & 31817.7  & 155              \\
\{EDF Energy, RWE Generation SE, SSE, Uniper UK Ltd\}                              & 38422.7  & 164              \\
\{EDF Energy, RWE Generation SE, SSE, Uniper UK Ltd, Scottish Power\}              & 42894.2  & 213              \\
\{EDF Energy, RWE Generation SE, SSE, Uniper UK Ltd, Scottish Power, Drax Power Ltd\} & 46929.2  & 216              \\ 
\bottomrule
\end{tabular}
\caption{Groups of GenCos that used bidding strategies, number of plants and total electricity generating capacity.}
\label{table:genco_table}
\end{table*}

For the reinforcement learning problem we have the following tuple: $(s_t,a_t,r_t,s_{t+1})$, where $(s_t, s_{t+1})$ is the state at time $t$ and $t+1$ respectively, $a_t$ is the action at time $t$ and $r_t$ is the reward at time $t$. For our problem the state space is given by the tuple shown in Equation \ref{eq:observation_tuple}:
\begin{equation}
\label{eq:observation_tuple}
	s_t=(H_i,D_i,p_{CO_{2}},p_{gas},p_{coal},p_{c}),
\end{equation}
\noindent where $H_i$ is the segment hour to bid into at timestep $i$, $D_i$ is the demand of the segment hour at timestep $t$, $p_{gas}$ is the price of gas, $p_{coal}$ the price of coal, $p_{CO_{2}}$ is the carbon tax price, and $p_{c}$ is the clearing price. We set the reward, $r_t$ to be the average electricity price of that time step, $p_{avg}$, as shown by Equation \ref{eq:reward}:

\begin{equation}
\label{eq:reward}
	r_t=p_{avg},
\end{equation}

\noindent where $p_{avg}$ is defined in Equation \ref{eq:p_avg}:

\begin{equation}
\label{eq:p_avg}
	p_{avg}=\frac{\sum_{m=0}^{M}p_m}{M},
\end{equation}

\noindent where $p_m$ is the price for the $m$ load segment for the respective time step, and $M$ is the total load segments, which in our case is 20. 

This reward function allowed us to closely match the profit that would be made by the GenCo. For instance, if a bid was not accepted, the GenCo would receive \textsterling0. If the bid was accepted, however, the total money received would be proportional to the reward, giving the GenCo an incentive to learn the dynamics of the electricity market clearing process. 

We set the replay buffer size, $\mathcal{R}$, to be 50,000. This enabled us to store a relatively large amount of history, whilst maintaining tractability.

For the action space, $a_t$, we modelled two scenarios, as shown by Equations \ref{eq:action-space-scen-1} and \ref{eq:action-space-scen-2}:

\begin{equation}
\label{eq:action-space-scen-1}
	0 \leq a_t \leq 600,
\end{equation}

\begin{equation}
\label{eq:action-space-scen-2}
	0 \leq a_t \leq 150,
\end{equation}

\noindent where there was a price cap of \textsterling$600$/MWh for Equation \ref{eq:action-space-scen-1} and \textsterling$150$/MWh for Equation \ref{eq:action-space-scen-2}. Only two values were chosen to reduce computational load. We chose \textsterling$150$/MWh as a reasonable price cap that may be introduced by a Government. This was roughly double the average accepted price in 2018, therefore allowed for higher prices in times of high demand or low supply. The \textsterling 600/MWh was chosen to simulate an unbounded price cap. This enabled us to see the price that an equilibrium is reached within a market with agents with market power.


In this work, we assume that the GenCo groups have no information about the generation capacity, marginal cost, bid prices or profits of other GenCos \cite{EsmaeiliAliabadi2017}. They learn the maximum profit that can be made through experience within a particular market. We assume this because in real-life GenCo groups have little information on their competitors sensitive bidding data. If they were to have perfect information on all their competitors they would be able to devise a perfect strategy which would always maximise their profit.



Note that in the model, the GenCo does not take any strategic action, that is, the GenCo is oblivious to the actions of other GenCos explicitly in its decision process. In fact, it does not have information on other GenCos. The GenCo is modelled as a simple agent that learns only from its own experience. GenCos' collective behaviour, however, may lead to strategic outcomes.\cite{EsmaeiliAliabadi2017}\\

\section{Results}
\label{sec:results}

In this section, we detail the results of the RL algorithm, and the effect that capacity has on average electricity price within the UK. Our approach could be generalised to any other decentralised electricity market in other countries. 

\begin{figure}
	\centering
    \includegraphics[width=0.49\textwidth]{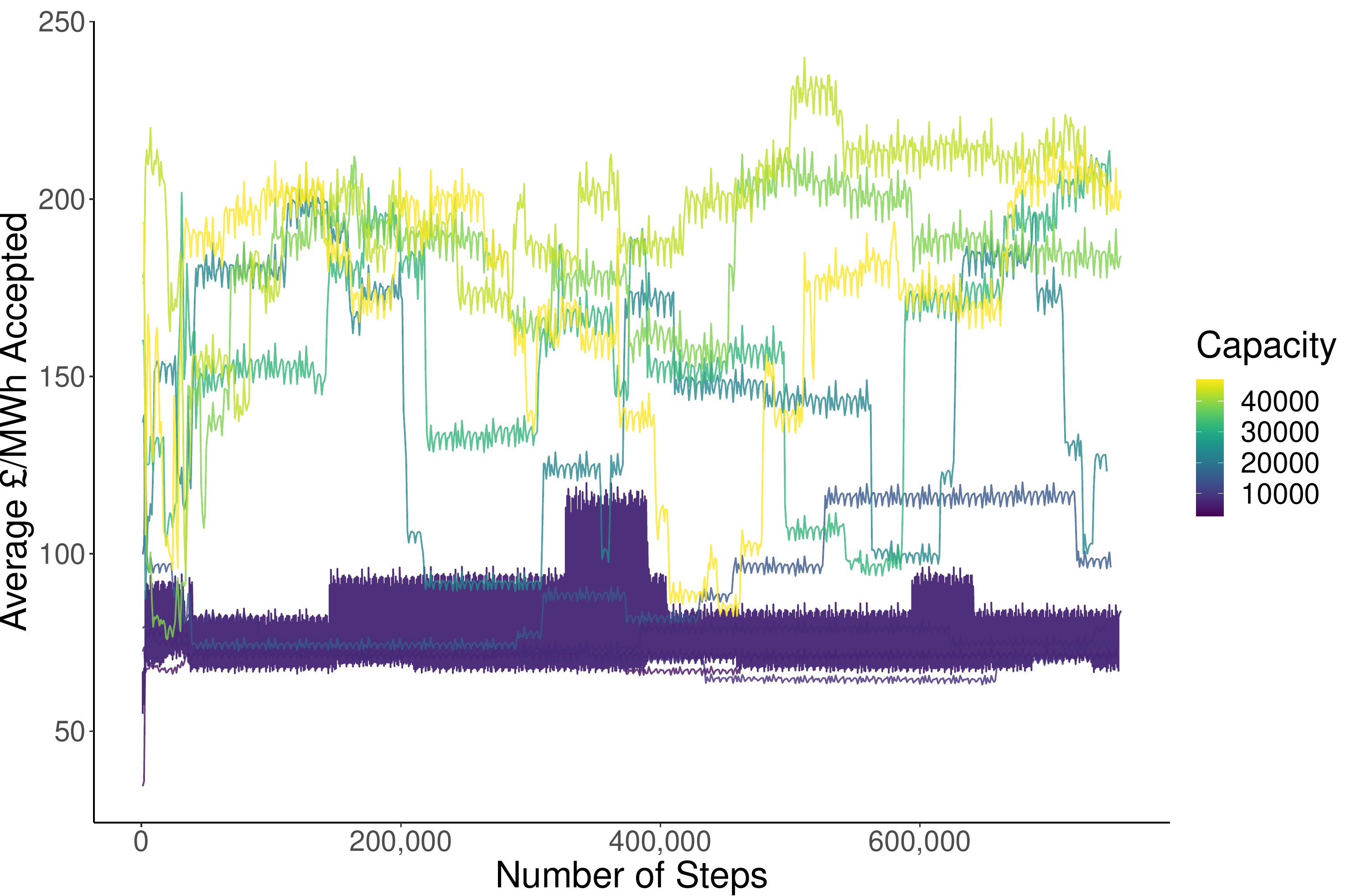}
    \caption{Reward over time for different groups of GenCos, max bid = \textsterling $600$/MWh.}
    \label{fig:unbounded_timesteps}
\end{figure}

\begin{figure}
	\centering
    \includegraphics[width=0.49\textwidth]{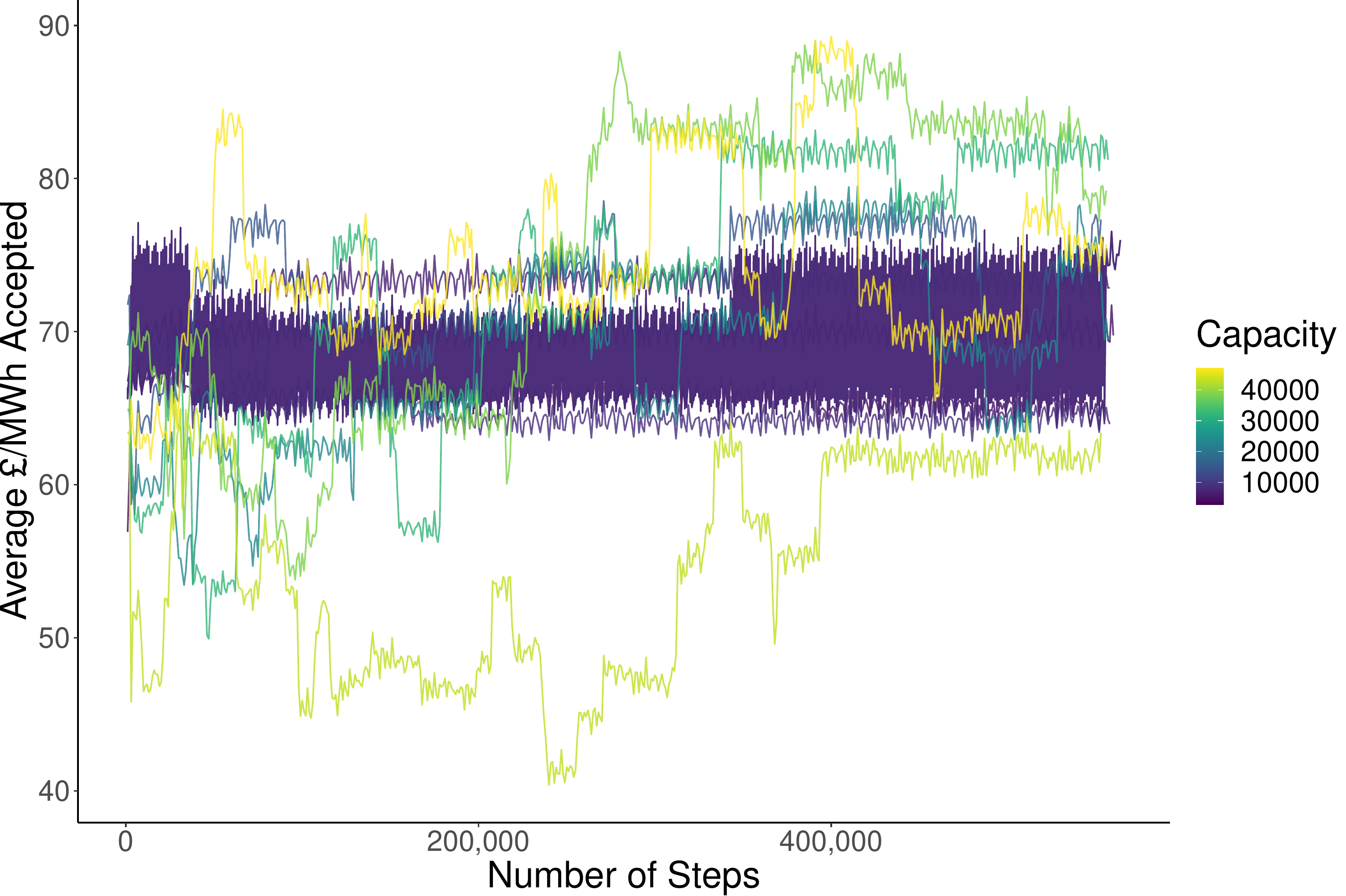}
    \caption{Reward over time for different groups of GenCos, max bid = \textsterling $150$/MWh.}
    \label{fig:bounded_timesteps}
\end{figure}

Figures \ref{fig:unbounded_timesteps} and \ref{fig:bounded_timesteps} show the rewards over a number of time steps for the unbounded and bounded cases respectively. Figure \ref{fig:unbounded_timesteps} shows a clear difference between agents which use the DDPG RL strategy and have a large capacity (green and yellow) compared to those which have a smaller capacity (dark purple). The axis in Figure \ref{fig:unbounded_timesteps} are much larger than those of \ref{fig:bounded_timesteps}, highlighting the effect of market power on an unbounded market.

The average electricity price for a capacity below 30,000MW, or ${\sim35\%}$ of total capacity, remains stable between \textsterling70/MWh and \textsterling100/MWh. This range may be due to the stochasticity in calculating the weights for the DDPG algorithm. The average electricity price does not change over the time steps or training. We, therefore, hypothesize that there is no market power as long as an individual GenCo owns below ${\sim}35\%$ of total electrical capacity. 

On the other hand, once the capacity of a GenCo or groups of GenCos is above 30,000MW, there is a significant increase in the average price for capacity. The average electricity price for capacity falls between \textsterling$170$/MWh and \textsterling$220$/MWh. 

\begin{figure}[]
	\centering
    \includegraphics[width=0.49\textwidth]{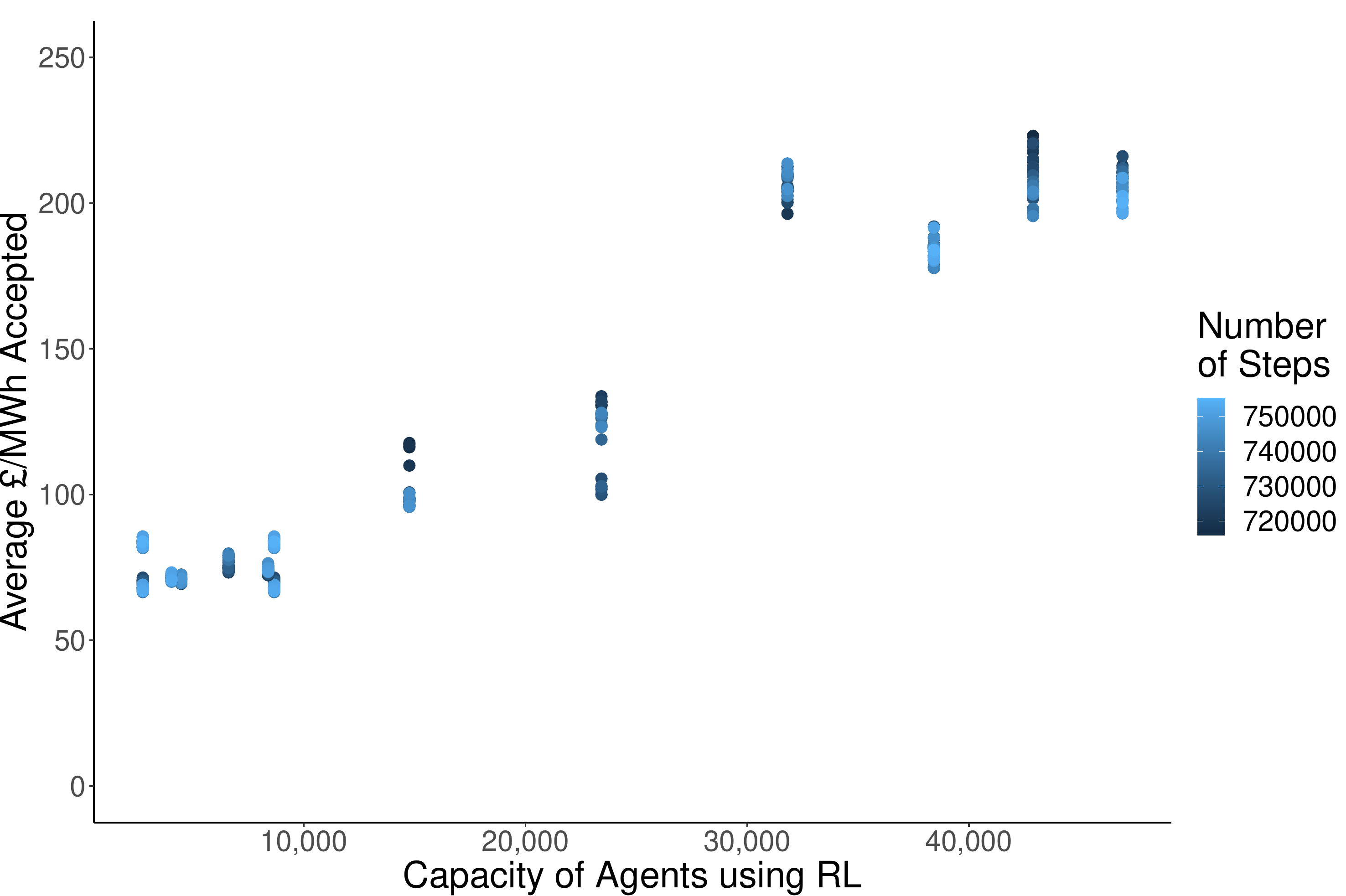}
    \caption{Capacity of agents using RL vs. average electricity price accepted, for unbounded agents.}
    \label{fig:unbounded_results_scatter}
\end{figure}

Figure \ref{fig:unbounded_results_scatter} displays the capacity controlled by the agents that use the RL strategy versus the average electricity price for the unbounded case. The color displays the number of steps. The step-change, as shown in Figure \ref{fig:unbounded_timesteps} can be seen clearly here, with agents with a capacity larger than ${\sim}$25,000MW causing a step change in electricity price. Electricity prices seems to cluster below ${\sim}$10,000MW. However, after this point, the average electricity price begins to increase.

\begin{figure}
	\centering
    \includegraphics[width=0.49\textwidth]{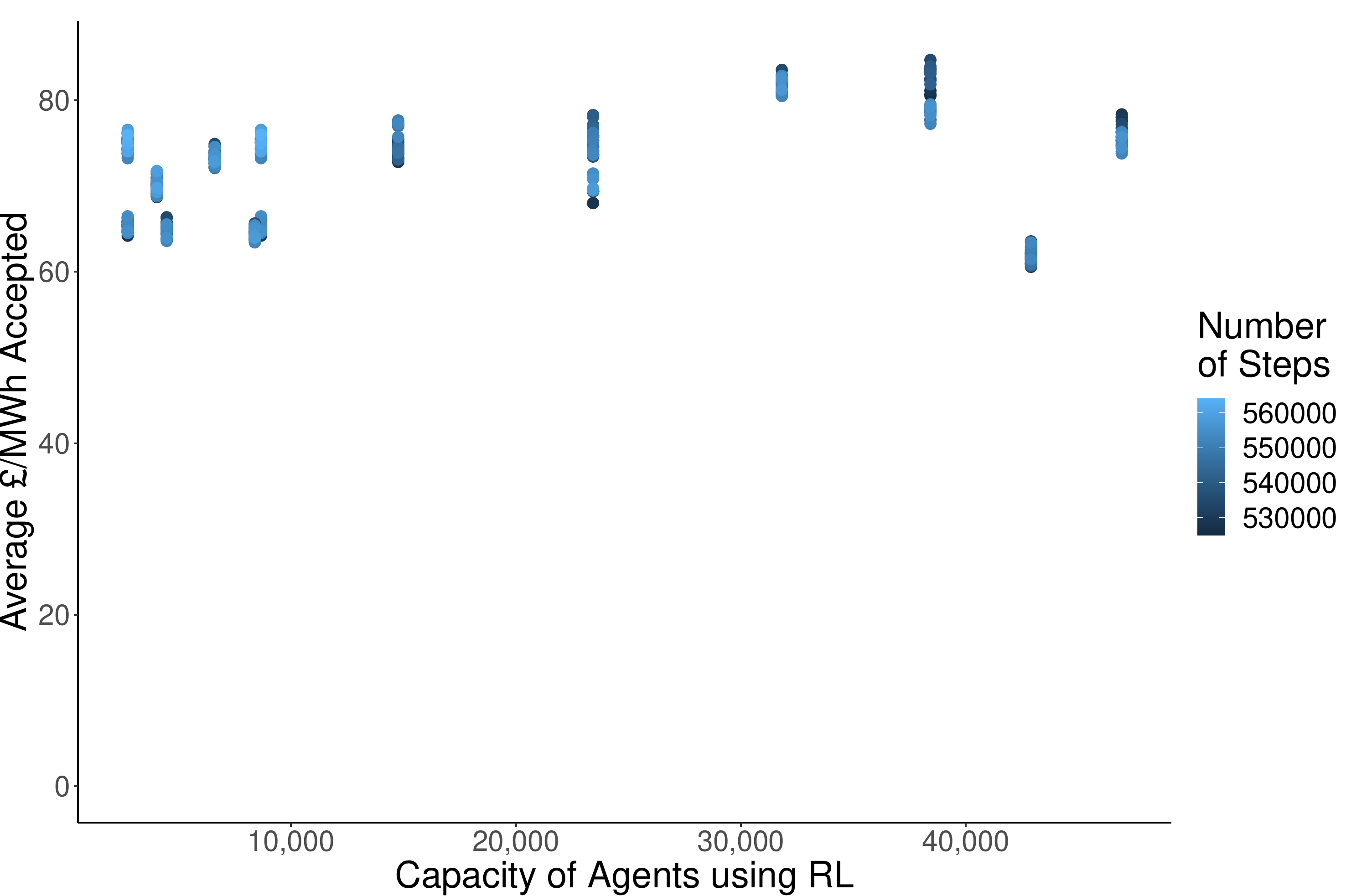}
    \caption{Capacity of agents using RL vs. average electricity price accepted, for bounded agents.}
    \label{fig:bounded_results_scatter}
\end{figure}

Figure \ref{fig:bounded_timesteps} shows a cluster between ${\sim}$\textsterling$60$/MWh and ${\sim}$\textsterling$80$/MWh irrespective of the capacity of the agents. This is verified by Figure \ref{fig:bounded_results_scatter}. This seems to suggest that setting a lower market cap reduces the ability for generators, irrespective of size, from influencing the electricity price.

\begin{figure}
\centering
\begin{subfigure}[b]{0.49\textwidth}   
\includegraphics[width=\columnwidth]{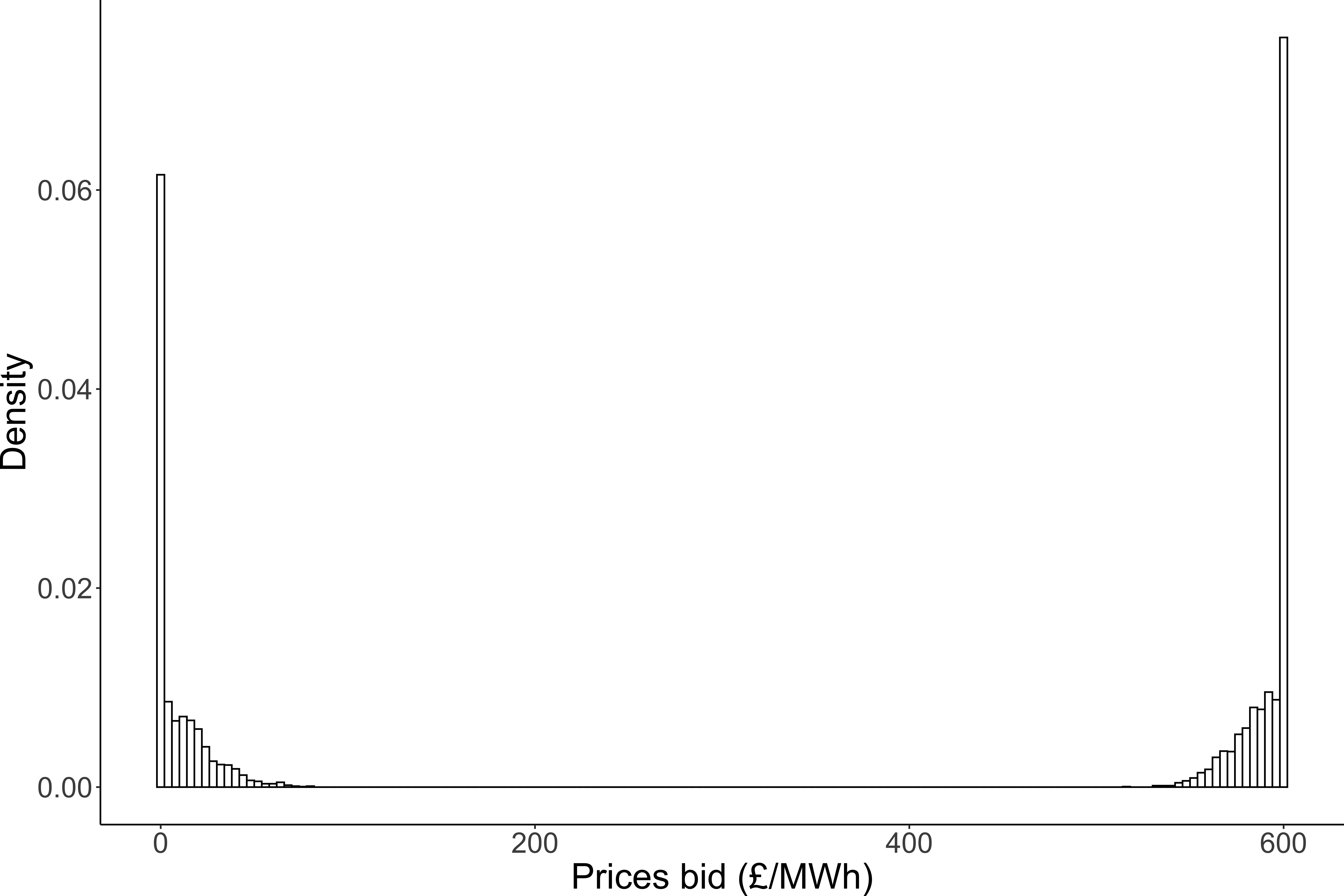}
\caption{Largest group of GenCos with a total controlled capacity of 46,929.2MW with a market cap of \textsterling600/MWh.}
\label{fig:large_company_capped_600_bids}
\end{subfigure}
\hfil
\begin{subfigure}[b]{0.49\textwidth}   
\includegraphics[width=\columnwidth]{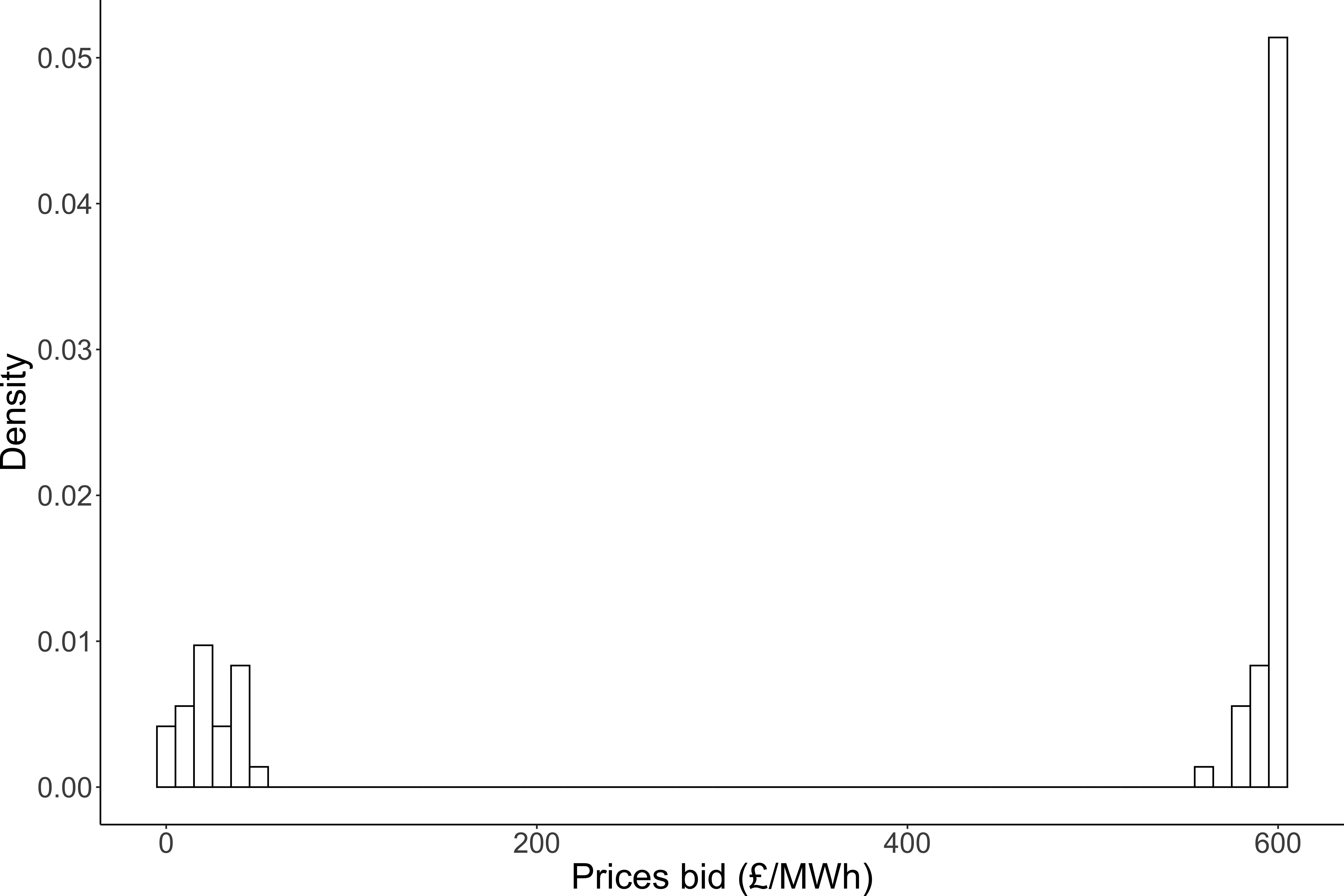}
\caption{Smallest company with a total controlled capacity of 2,738.7MW with a market cap of \textsterling600/MWh.}
\label{fig:small_company_capped_600_bids}
\end{subfigure}
\caption{Bids made by generator companies with a market cap of \textsterling600/MWh.}
\label{fig:capped_600_bids}
\end{figure}

\begin{figure}
\centering
\begin{subfigure}[b]{0.49\textwidth}   
\includegraphics[width=\columnwidth]{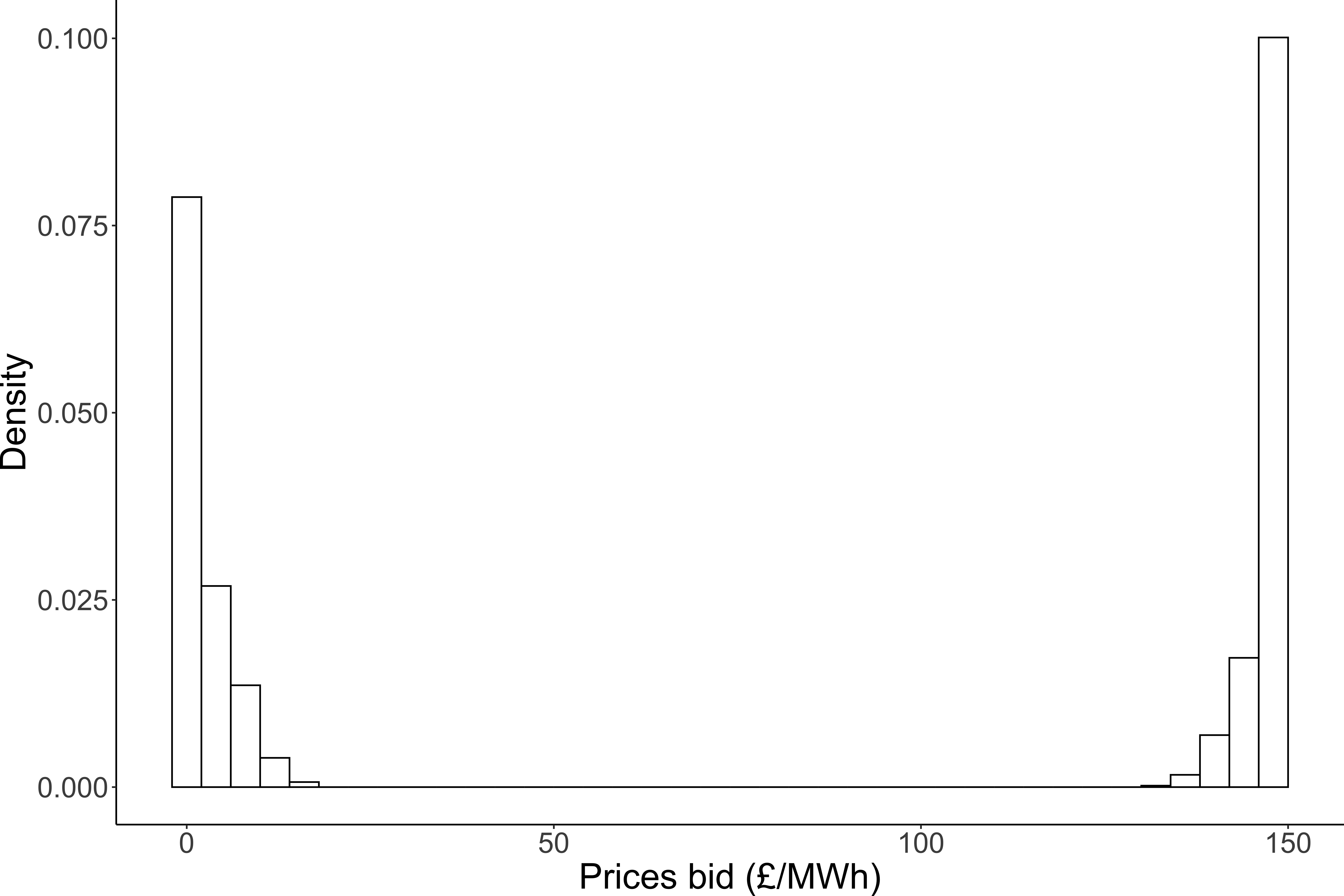}
\caption{Largest group of GenCos with a total controlled capacity of 46,929.2MW with a market cap of \textsterling150/MWh.}
\label{fig:large_company_capped_150_bids}
\end{subfigure}
\hfil
\begin{subfigure}[b]{0.49\textwidth}   
\includegraphics[width=\columnwidth]{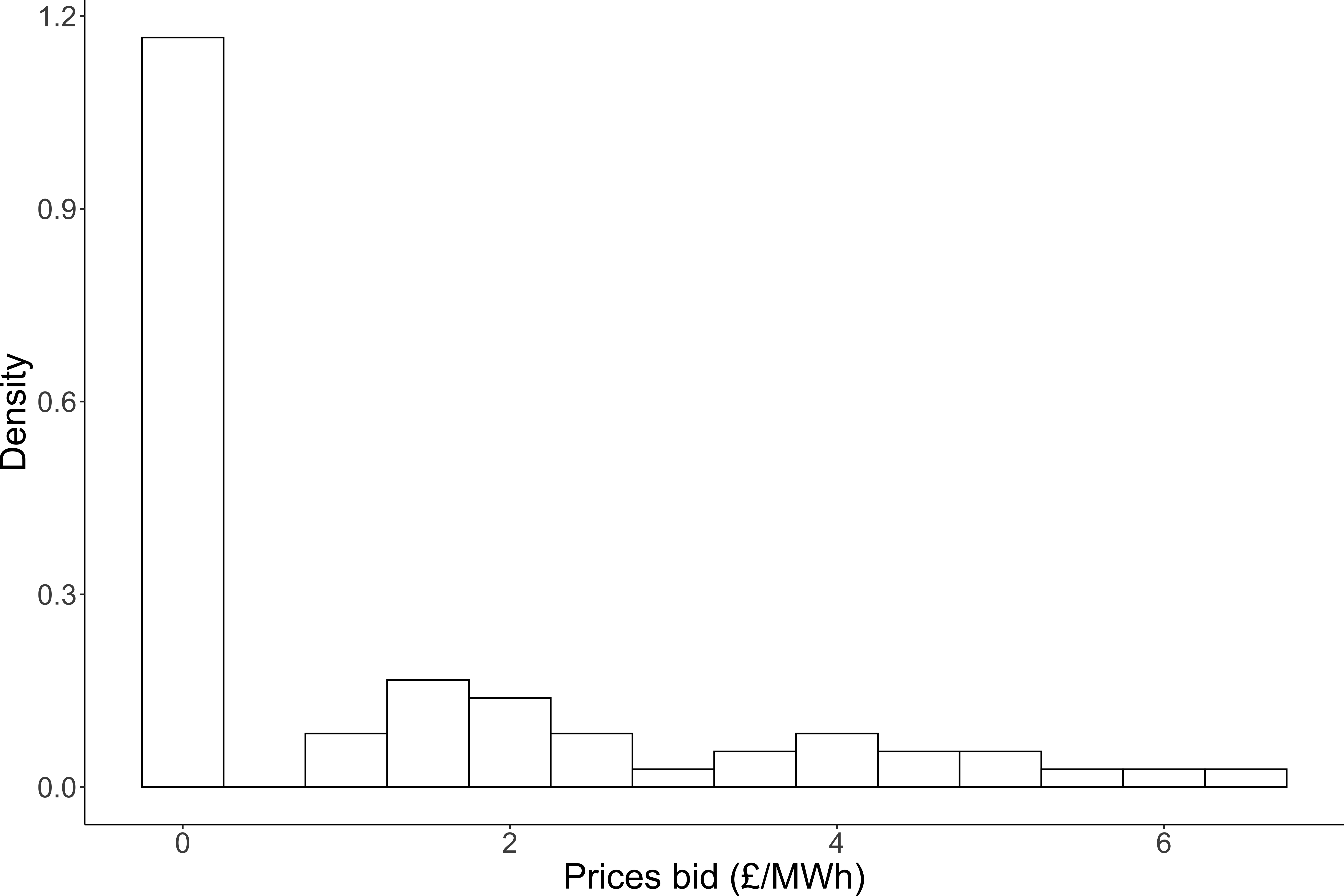}
\caption{Smallest company with a total controlled capacity of 2,738.7MW with a market cap of \textsterling150/MWh.}
\label{fig:small_company_capped_150_bids}
\end{subfigure}
\caption{Bids made by generator companies with a market cap of \textsterling150/MWh.}
\label{fig:capped_150_bids}
\end{figure}

Figures \ref{fig:capped_600_bids} and \ref{fig:capped_150_bids} display the bids which use the RL algorithm for every power plant at the end of training. For instance, in Figure \ref{fig:large_company_capped_600_bids}, the group with a capacity of 46,929.2, as shown by Table \ref{table:genco_table}, bids a majority of the time the maximum bid (\textsterling600). The second most common bid is \textsterling0, or the minimum allowed bid. The bids then exponentially fall around these two peaks.

The number of bids made by each GenCo changes dependent on the number of plants that they own, with the Orsted GenCo only making eleven bids per segment, and the largest group making 216 bids per segment (as shown by Table \ref{table:genco_table}). Figure \ref{fig:capped_600_bids} displays the uncapped scenario (\textsterling 600/MWh) and \ref{fig:capped_150_bids} displays the capped scenario (\textsterling 150/MWh).

Figure \ref{fig:large_company_capped_600_bids} shows the bids made by the largest group of GenCos as shown in Table \ref{table:genco_table}. A bimodal distribution can be seen, where the group of GenCos tend to bid either the maximum (\textsterling600) or the minimum (\textsterling0) bid. We hypothesize that they bid the maximum amount as this ensures that the market price is artificially raised, and that they are able to utilize their market power. The minimum price is bid the rest of the time to ensure that generators bids are always accepted, regardless of whether the market price has been artificially raised or not in each respective clearing segment.

Figure \ref{fig:small_company_capped_600_bids} displays the bids of the small company with a market cap of \textsterling 600/MWh. The small company also seems to have a bimodal distribution; however, it bids the highest price more often (\textsterling600). This may be due to the fact that it is able to influence the price at certain market segments, and the reward of the higher accepted reward outweighs the times in which it is not accepted on the market segments. Again, bidding low seems to be the strategy in which to take if the GenCo does not believe it will be able to influence the final price. As the market simulated is a uniform pricing market, having a \textsterling0 bid accepted does not mean that the GenCo will be paid \textsterling0. Rather, the GenCo will be paid the market clearing segment, which is set by the most expensive power plant accepted onto that market segment. This is the basis for a uniform pricing market, a mechanism which is often used in electricity markets. We used this mechanism in our simulation.

Figures \ref{fig:large_company_capped_150_bids} shows the bids made by the largest group of GenCos in each market segment. It seems to take a similar strategy to that of the large company in the uncapped scenario, as shown in Figure \ref{fig:large_company_capped_600_bids}. We believe, as similar in the uncapped scenario, that this is due to the market power that this group possesses. Being able to influence the market price enables the GenCo group to inflate the prices. It also takes a conservative strategy to bid the minimum price allowed, to ensure that the bid is accepted regardless of price.

Figure \ref{fig:small_company_capped_150_bids} displays the strategy of the smallest company in the capped scenario. Here, it seems that the GenCo is unable to influence the price at all, and therefore bids \textsterling0/MWh for the majority of the time. This is similar to the expected strategy of GenCos, who tend to bid their short-run marginal cost to ensure that they do not miss out on potential profit. The short-run marginal cost can often change based upon fuel, carbon and generator type. However, for renewable energy it is near \textsterling0 and for fossil-fuel based plants it is near the cost of fuel and carbon at that point in time.

\begin{figure}
	\centering
    \includegraphics[width=0.49\textwidth]{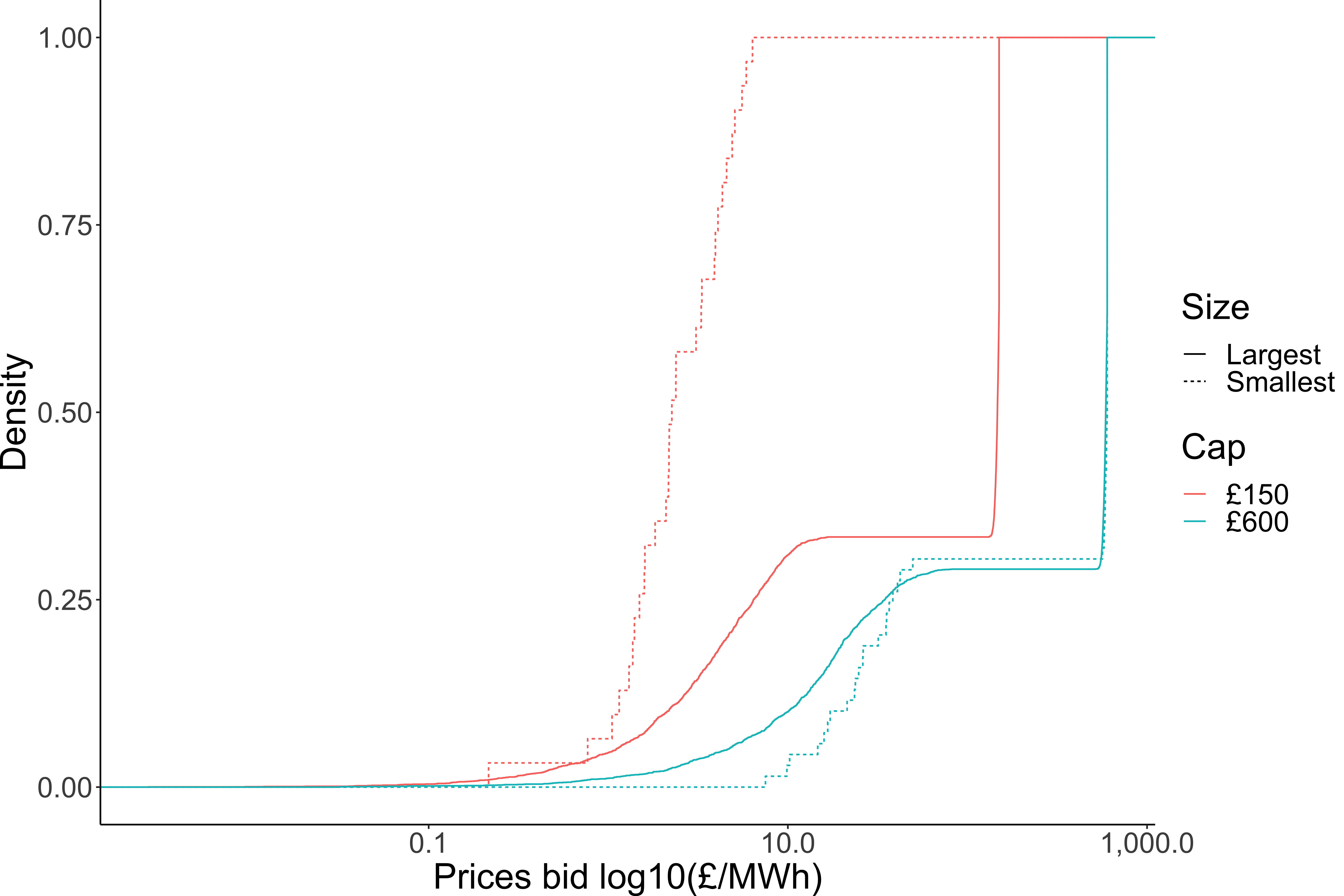}
    \caption{Empirical cumulative distribution function comparing the likelihood of each bid for the different sized GenCos and market cap.}
    \label{fig:ecdf_all}
\end{figure}

Figure \ref{fig:ecdf_all} displays an empirical cumulative distribution function, where we compare each of the scenarios to the bids actually made. The price cap of \textsterling600, shows similar results for both sizes of companies. However, the smaller company bids higher more often. With a price cap of \textsterling150, the smallest company is much more willing to bid a lower amount than the larger company, for reasons described previously.

\begin{figure}
    \includegraphics[width=0.49\textwidth]{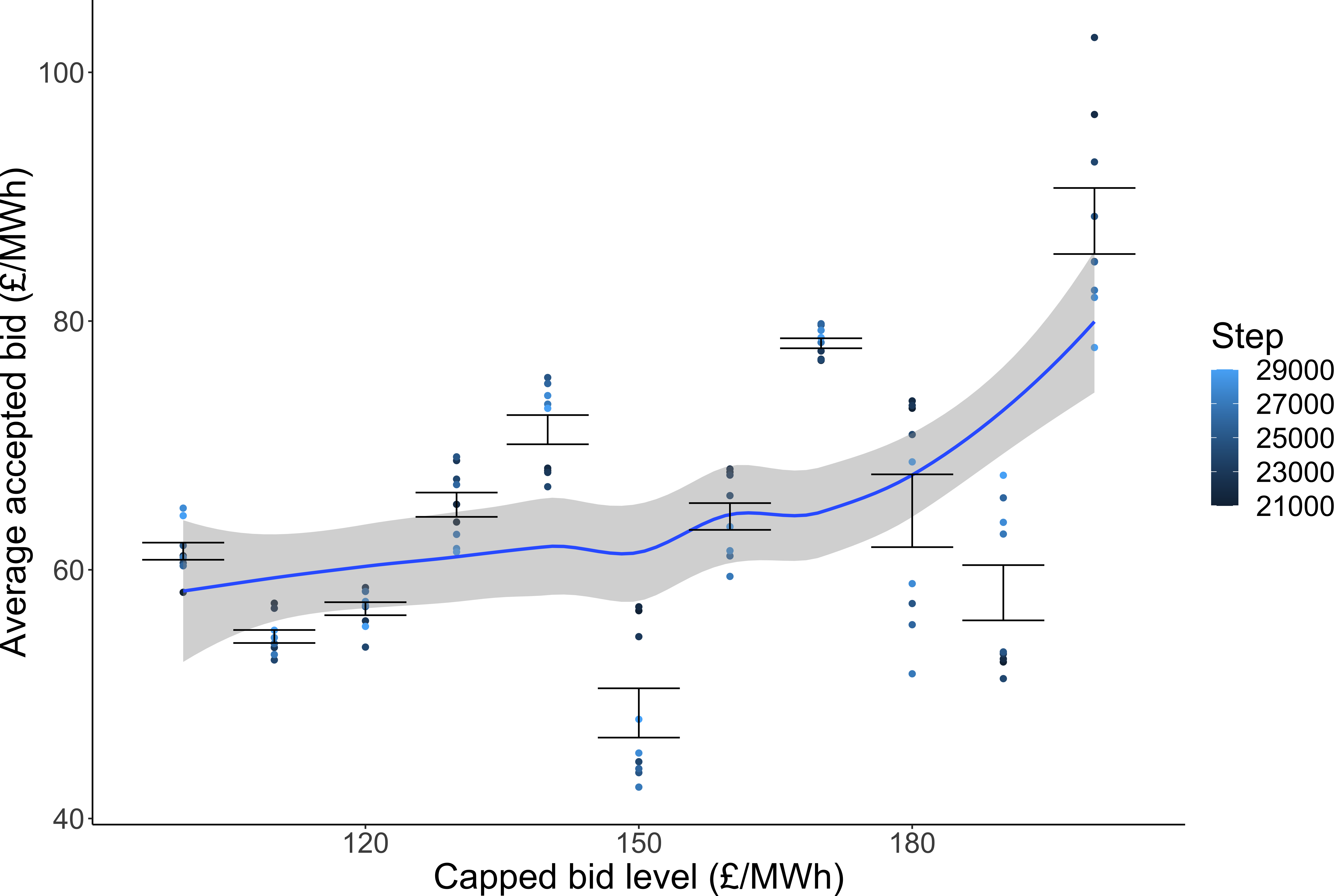}
    \caption{Capacity of agents using RL vs. average electricity price accepted, for bounded agents.}
    \label{fig:sensitivity_analysis}
\end{figure}

We ran a sensitivity analysis to observe the effects of the market cap on final average accepted bid price. The largest GenCo group used a strategy for this sensitivity analysis. Figure \ref{fig:sensitivity_analysis} displays the results. It seems that whilst the average accepted bid price increased with the capped bid level; there is a significant increase after a market cap of \textsterling190/MWh. This may be due to the case that the GenCos begin to outbid the SRMC bidding GenCos at this price point.


\section{Discussion}
\label{sec:discussion}

Our results demonstrate the ability for GenCos to artificially increase the electricity price through market power in an uncapped market. Our results have shown that in an uncapped market, any single agent or groups of agents who make bids using the same strategy and information, should have less than ${\sim}$10,000MW of controlled capacity. This defines the optimal capacity by any single GenCo to have a fair level of competition. After this, the electricity price begins to rise with no increase in outcome and welfare. It is also worth nothing that when the market is capped, the average accepted price does not simply become the capped price.

However, if there is an electricity market with a few large or colluding players, it is possible to remove their advantage through the introduction of a price cap. Our results show that whilst average accepted bids increase with market cap level, the value does not increase significantly until \textsterling$200$ is reached.


This information and approach can help to inform government policy to ensure fair competition within electricity markets, as well as run the model for their own scenario. It is hypothesized that the findings in this paper are generalizable to other decentralized electricity markets in other geographies due to their similar market structures. Whilst the figures presented here may not be the same; we hypothesize that the region of interest will be similar.


\section{Conclusion}
\label{sec:conclusion}

In this work, we used the deep deterministic policy gradient (DDPG) reinforcement learning method to make strategic bids within an electricity market. We utilized the agent-based model ElecSim to model the UK electricity market. We utilized the DDPG algorithm only for a certain subset of agents, from small individual generation companies (GenCos) to large groups of GenCos. 

This enabled us to explore the ability for GenCos with a large capacity to artificially increase the price in the electricity market within the UK if they are in control of a sufficiently large generation capacity. Our results show that the optimum level of control of any single GenCo or groups of GenCo is below ${\sim}$10,000MW or ${\sim}$11\% of the total capacity. Above this, prices begin to increase with no real additional benefit to the consumer. After ${\sim}$25,000MW, or ${\sim}$35\% of the total capacity, the prices begin to increase substantially, to ${\sim}$\textsterling200, over triple the original cost without this market power. The introduction of a market cap of \textsterling$150$ reduces all market power and maintains electricity price at a reasonable level.

We found through a sensitivity analysis, that the average electricity price in the market over a year remains low with a price cap smaller than \textsterling 190. However, after this level, the average electricity price begins to increase.

The bidding strategies of the GenCos which used the reinforcement learning algorithm was to bid both the maximum and minimum price possible. However, this was only in the case of GenCos, or groups of GenCos, which have a large controlled capacity without a market cap. If the strategically bidding GenCo had a relatively smaller capacity, then a market cap had a great effect on ensuring that these GenCos bid close to their short-run marginal cost. 

Our work has shown the ability for reinforcement learning to learn an optimal bidding strategy to maximize a GenCo's profit within an electricity market. The ability for GenCos to use their market power is also highlighted, and is dependent on electricity generation capacity of the respective GenCo.

The information presented here can be used by governments of liberalised electricity markets to ensure fair competition between generator GenCos. For instance, governments could ensure a wide range of participants have an equally distributed proportion of generation capacity. If this is not possible within such a market, it is possible for the introduction of market caps to limit the impact that a few large GenCos have on the market. Whilst the numbers presented in this work may not be directly translatable to other markets, the methodology presented here is generalisable.

In future work, we would like to enable GenCos to withhold the capacity on offer to the electricity market. This would enable further market power by reducing competition further.  Additionally, we would like to assess the market power in different countries with different market structures and total electricity supply.


\section{Acknowledgment}

This work was supported by the Engineering and Physical Sciences Research Council, Centre for Doctoral Training in Cloud Computing for Big Data [grant number EP/L015358/1].


\bibliographystyle{IEEEtran}
\bibliography{library,custom_bib,custombibtexfile}

\end{document}